\documentclass{nature2}
\usepackage{graphicx}
\usepackage[usenames]{color}
\usepackage{xcolor}
\usepackage[export]{adjustbox}
\usepackage{amsmath}
\usepackage{amssymb}
\usepackage{chemformula,array}

\title{Enhanced elastic stability of a topologically disordered\\ crystalline metal--organic framework}

\author{Emily G. Meekel,$^1$ Phillippa Partridge,$^2$ Robert A. I. Paraoan,$^1$ Joshua J. B. Levinsky,$^2$ Ben Slater,$^3$ Claire L. Hobday$^2$ \& Andrew L. Goodwin$^{1\ast}$}

\begin{document}

\maketitle

\begin{affiliations}
	\item Inorganic Chemistry Laboratory, Department of Chemistry, University of Oxford, Oxford, UK
	\item Centre for Science at Extreme Conditions and EaStCHEM School of Chemistry, University of Edinburgh, Edinburgh, UK
	\item Department of Chemistry, University College London, London, UK
\end{affiliations}

\begin{abstract}

By virtue of their open network structures and low densities, metal--organic frameworks (MOFs) are soft materials that exhibit elastic instabilities at low applied stresses. The conventional strategy for improving elastic stability is to increase the connectivity of the underlying MOF network, which necessarily increases material density and reduces porosity. Here we demonstrate an alternative paradigm, whereby elastic stability is enhanced in a MOF with an aperiodic network topology. We use a combination of variable-pressure single-crystal X-ray diffraction measurements and coarse-grained lattice-dynamical calculations to interrogate the high-pressure behaviour of the topologically aperiodic system TRUMOF-1, which we compare against that of its ordered congener MOF-5. We show that the topology of the former quenches the elastic instability responsible for pressure-induced framework collapse in the latter, much as irregularity in the shapes and sizes of stones acts to prevent cooperative mechanical failure in drystone walls. Our results establish aperiodicity as a counterintuitive design motif in engineering the mechanical properties of framework structures, relevant to MOFs and larger-scale architectures alike.

\end{abstract}

\section*{Introduction}

Metal--organic frameworks (MOFs) are porous crystalline materials, the scaffolding-like structures of which are assembled from inorganic nodes connected by organic linkers.\cite{Li_1999,Eddaoudi_2000} Varying the geometry and connectivity of node and linker components allows access to an enormous variety of different network topologies.\cite{OKeeffe_2014} From a mechanical perspective, MOFs are significantly more elastically compliant than conventional inorganic solids,\cite{Horike_2009,Tan_2011,Coudert_2015} with moduli approaching those typical of organic polymers such as rubber.\cite{Tan_2011} This mechanical softness is primarily a consequence of the open network architecture of MOFs and the flexibility of their structural elements;\cite{Ortiz_2012} it is also a key ingredient in driving anomalous mechanical phenomena such as negative linear compressibility\cite{Li_2012,Cairns_2015} and negative gas adsorption.\cite{Krause_2016} However, the low elastic moduli of their structures mean that, when evacuated, MOFs are readily susceptible to collapse under applied pressure, exhibiting pressure-induced amorphisation and/or phase transitions to dense polymorphs.\cite{Collings_2019} Such low-pressure collapse is an important consideration for practical applications, since many industrial processes involve stresses greater than the typical elastic stability limits of MOFs.\cite{Chapman_2009} The critical pressure at which deformation occurs---itself closely related to the magnitude of the elastic moduli\cite{Mouhat_2014}---scales loosely with network connectivity and strength of metal--linker interactions.\cite{Wu_2013,Banlusan_2017} For example, the four-connected zeolitic imidazolate framework ZIF-8 (Zn$^{2+}$--N links) amorphises at 0.34\,GPa whereas the twelve-connected framework UiO-66 (Zr$^{4+}$--O links) maintains crystallinity up to a threshold value of 1.4\,GPa;\cite{Chapman_2009,Bennett_2016,Yot_2016} the corresponding shear moduli are 0.97 and 14\,GPa, respectively.\cite{Tan_2012,Wu_2013} Hence the conventional design focus for strengthening MOFs has centred primarily on high-connectivity network structures assembled from highly-charged cations.\cite{Moghadam_2019}

In entirely different contexts, it is well appreciated that irregular architectures can confer mechanical strength through frustration of collapse mechanisms.\cite{Qin_2013,Fernandes_2021,Liu_2022,Yang_2022}  An archetypal example is that of varying stone size and shape to strengthen `drystone wall' structures; another is the complex disordered channel networks in termite nests, which optimise structural stability for given resource and are the inspiration for a new class of advanced metamaterials.\cite{Reid_2018,Heyde_2021,Liu_2022,Senhora_2022} An obvious question is whether these same ideas might be relevant on the atomic scale to improve the elastic stability of functional materials such as MOFs.

\begin{figure}
	\begin{center} 
		\includegraphics{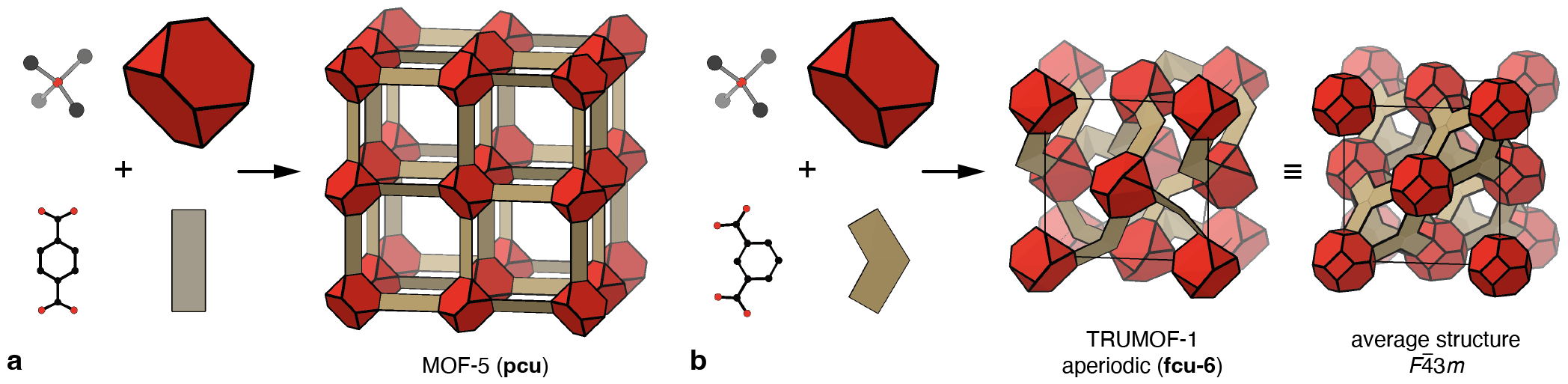}
	\end{center}
	\caption{\label{fig1}\footnotesize{\bf Assembly of MOF-5 and TRUMOF-1 structures from building units.} {\bf a} In MOF-5, tetrahedral OZn$_4$ clusters (red polyhedra) are connected by 1,4-bdc linkers (beige panels) to form a network structure with the simple-cubic {\bf pcu} topology. {\bf b} The structure of TRUMOF-1 is generated by similar components, except that the linear 1,4-bdc linker is replaced by the bent 1,3-bdc isomer (beige angled panels). Clusters are arranged on a face-centred cubic lattice, with each cluster connected to six of its twelve nearest neighbours (we call this topology {\bf fcu-6}). The connectivity in TRUMOF-1 is aperiodic; shown in the centre of this panel is a representation of one possible $1\times1\times1$ approximant. The configurational average of the TRUMOF-1 structure, which is periodic but with partial site occupancies, is represented on the right-hand side.} 
\end{figure}

We have recently discovered the material TRUMOF-1, a crystalline MOF whose structure is based on an unusual aperiodic network topology related to so-called Truchet tilings.\cite{Meekel_2023,Smith_1987} The chemistry of TRUMOF-1 is almost identical to that of the canonical system MOF-5: both are assembled from octahedrally-coordinated OZn$_4$ clusters connected by benzenedicarboxylate (bdc) linkers to form a network of uniformly six-connected nodes [Fig.~\ref{fig1}].\cite{Li_1999} In the case of MOF-5, the linear connectivity of 1,4-bdc links nodes to give a network with the simple cubic (\textbf{pcu}) topology. The nodes of TRUMOF-1 are also six-connected with local octahedral geometries, but the network formed by the bent 1,3-bdc linkers is aperiodic and never repeats.\cite{Meekel_2023} The resulting connectivity is based on non-random partial occupancy of the 12-connected \textbf{fcu} net in a way that is locally homogeneous but not long-range ordered; in the absence of a formal language for such aperiodic networks, we use the symbol \textbf{fcu-6} to denote this aperiodic six-connected variant of the \textbf{fcu} topology. The key point is that the two materials share the same degree of network connectivity, and the same node--linker chemistry, but by virtue of a subtle change in linker geometry differ in terms of the presence or absence of topological periodicity.

Here we report the use of variable-pressure single-crystal X-ray diffraction measurements to study the elastic behaviour of TRUMOF-1 under hydrostatic compression.\cite{McKellar_2015} Whereas evacuated MOF-5 is famously unstable under pressure (it amorphises at $p\lesssim0.2$\,GPa, Refs.~\citenum{Hu_2010,Erkartal_2018,Pallach_2021,Baxter_2022}), we find that TRUMOF-1 remains crystalline to the hydrostatic limit (1.8\,GPa) of a suitable non-penetrating pressure-transmitting medium (PTM). Indeed, our experimental measurements establish TRUMOF-1 as the most elastically stable isotropic MOF reported to date.\cite{anisotropynote} Using a coarse-grained lattice-dynamical model that captures the key mechanical effects of topological aperiodicity in TRUMOF-1, we rationalise the enhancement of elastic stability of this system relative to MOF-5, and demonstrate that its compression mechanism involves activation of internal degrees of freedom in a way that varies spatially throughout the TRUMOF-1 structure. We argue that these displacements provide a shock-absorption mechanism that is closely related to both the phenomenon of combinatorial mechanics proposed in Ref.~\citenum{Reynolds_2021} and the spatially-inhomogeneous mechanical responses identified in disordered metamaterials.\cite{Horrigan_2009,Liu_2022} Our study suggests how aperiodic network architectures of the kind adopted by TRUMOF-1 might be exploited in the rational design of low-density framework structures with useful mechanical properties.

\section*{Results}

\begin{figure}
	\begin{center} 
		\includegraphics[width=\textwidth]{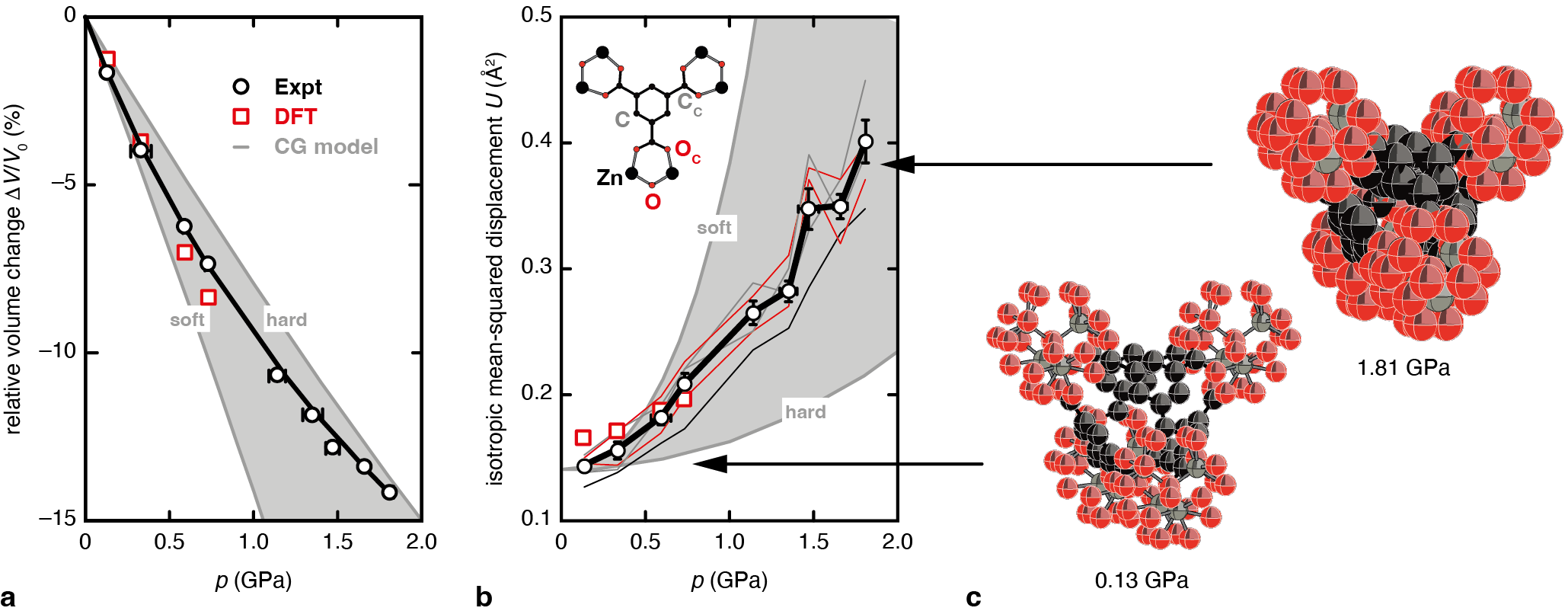}
	\end{center}
	\caption{\label{fig2}\footnotesize{\bf Structural response of TRUMOF-1 to hydrostatic pressure.} {\bf a} The $V(p)$ equation of state obtained using single-crystal X-ray diffraction measurements (open circles) is well accounted for by a third-order Birch--Murnaghan fit (solid line). The low-pressure behaviour obtained from DFT-driven geometry optimisations of a $2\times2\times2$ approximant is shown as open squares and that of our coarse-grained lattice-dynamical models (``CG model'') is shown as a shaded region. {\bf b} The isotropic mean-squared displacement magnitudes extracted from our crystallographic analysis increase with pressure; trends for individual atom types are shown as thin lines, with the average shown as open circles and bold line. Estimated values extracted from our DFT configurations are shown as open squares. The CG and DFT data have been shifted vertically by 0.14\,\AA$^2$ to account for the thermal contribution to $U$. {\bf c} Representations of fragments of the TRUMOF-1 structure at low and high pressure, with displacement ellipsoids shown at 50\% probability. Zn, O, and C atoms shown in grey, red, and black, respectively.} 
\end{figure}

\noindent{\bf High-pressure crystallography.} The variable-pressure X-ray diffraction behaviour of evacuated TRUMOF-1 single crystals is relatively straightforward: throughout the hydrostatic regime of the non-penetrating PTM used in our study (DAPHNE oil 7373;\cite{Yokogawa_2007} 0--1.8\,GPa) we observed pressure to effect a smooth and monotonic decrease in size of the cubic $F\bar43m$ unit-cell of TRUMOF-1 [Fig.~\ref{fig2}(a)]. In particular, there was no evidence for any symmetry-lowering phase transitions from the ambient-pressure phase, nor was there any indication of pressure-induced amorphisation. As straightforward as this behaviour may seem, it is nonetheless surprising: our results show that TRUMOF-1 outperforms all known isotropic MOFs in terms of stability under hydrostatic compression.\cite{Coudert_2015,McKellar_2015} This resilience to applied pressure is not a consequence of TRUMOF-1 being particularly stiff. The experimental $V(p)$ equation of state corresponds to a zero-pressure bulk modulus $B_0=7.5\pm0.5$\,GPa, which is much lower than that of MOF-5 (15--40\,GPa, Refs.~\citenum{Banlusan_2017,Rogge_2018,Baxter_2022}). The pressure derivative of the bulk modulus, determined using a third-order Birch--Murnaghan fit to our data, was $B^\prime=5.9(7)$. Density functional theory (DFT) calculations carried out for a set of ten $1\times1\times1$ approximants\cite{Thygesen_2017,Meekel_2023} of the TRUMOF-1 structure gave values of the zero-pressure bulk modulus across the range $4.8\leq B_0\leq7.4$\,GPa, the larger values correlating with denser approximant structures. Geometry optimisation of a larger set of approximants (\emph{viz.}\ $2\times2\times2$) at a series of small applied pressures gives volume reductions that are close to our experimental values (Fig.~\ref{fig2}(a); see SI for further discussion). Mindful of the complexity of TRUMOF-1, we consider this agreement between DFT and experiment to be remarkably good. So our first key finding is that TRUMOF-1 is a compliant material that is nonetheless able to accommodate compression to a significantly larger degree than other, stiffer, MOFs.

Our diffraction data were sufficiently complete to allow crystal-structure refinement as a function of applied pressure. These refinements considered only the Bragg component to the diffraction pattern (the diffuse scattering interpreted in Ref.~\citenum{Meekel_2023} being too weak for us to measure accurately in a pressure cell). We found no meaningful change to the average structure of TRUMOF-1, other than a strong increase in the magnitude of atomic displacement parameters with increasing pressure [Fig.~\ref{fig2}(b,c)]. Analysis of residual electron density in Fourier maps showed no appreciable variation in pore content with pressure (see SI), which is consistent with the exclusion of PTM from the pore network throughout our measurements.\cite{Yokogawa_2007,Baxter_2022} Hence our results can be interpreted in terms of the intrinsic behaviour of TRUMOF-1 itself, free from guest-inclusion effects.\cite{Graham_2011,Baxter_2022,Collings_2019}

The increase in magnitude of atomic displacements with pressure is at once both anomalous and counterintuitive. After all, densification usually dampens vibrational motion such that any measure of atomic displacements decreases accordingly as pressure is increased.\cite{Anderson_2001} This effect is captured formally by the Gr{\"u}neisen relation $\gamma = -\partial\ln\omega/\partial\ln V$ which links changes in volume $V$ to changes in phonon frequencies $\omega$.\cite{Gruneisen_1926} In most materials, $\gamma\simeq1$, and hence phonons stiffen under pressure, reducing the amplitude of thermal motion. Exceptions are known, and perhaps the best-studied are those of phonon-driven negative thermal expansion (NTE) materials such as ScF$_3$ and Zn(CN)$_2$,\cite{Greve_2010,Williams_1997} for which $\gamma<0$ and the amplitude of volume-reducing phonon modes actually increases as volume is reduced.\cite{Wei_2020,Fang_2013} TRUMOF-1, however, does not show NTE,\cite{Meekel_2023} and we will come to rationalise the anomalous increase in displacement magnitude in terms of an alternative mechanism whereby pressure magnifies the degree of static disorder.

\noindent{\bf Coarse-grained lattice dynamical model.} At face value, there is no obvious reason why TRUMOF-1 should be so much more elastically stable than MOF-5---after all, the two systems share the same chemistry and the same degree of network connectivity. In order to understand how their topological differences result in such contrasting mechanical responses, we sought to develop the simplest possible lattice-dynamical model capable of capturing the key elastic properties of the two systems. Our approach (inspired by that used elsewhere\cite{Dove_2020} for ScF$_3$) was as follows. First we developed a simplified model of MOF-5 in which OZn$_4$ clusters were mapped onto individual sites, and linkers were mapped onto harmonic springs connecting one site to its six nearest neighbours. Additional harmonic (three-body) springs were included amongst neighbouring triplets to capture the angular rigidity of the network. In this way, the lattice energy of MOF-5 is coarse grained as follows:
\begin{equation}\label{hamil1}
    E_{\rm latt}=\frac{1}{2}k_{r}\sum_{i>j}(r_{ij}-r_{\rm e})^2 + \frac{1}{2}k_\theta\sum_{i>j>k}(\theta_{ijk}-\theta_{\rm e})^2.
\end{equation}
Here, $k_r$ and $k_\theta$ are effective force constants that capture the resistance of MOF-5 to linear and shear deformations, respectively. The sums in Eq.~\eqref{hamil1} are taken over connected nodes $i,j(,k)$ and $r_{\rm e},\theta_{\rm e}$ represent the equilibrium node separation (= half the MOF-5 unit-cell parameter) and the relevant intra-framework angle (= $90^\circ$ or $180^\circ$ as appropriate). The two free parameters of this model---namely the numerical values of $k_r$ and $k_\theta$---were fixed so as to reproduce the $C_{11}$ and $C_{44}$ elastic constants of MOF-5. As it happens, there is a large spread amongst measures of these values reported from both experimental and computational studies,\cite{Zhou_2006,Bahr_2007,Banlusan_2017,Yan_2023} so we consider limiting cases of both `soft' and `hard' parameterisations (see SI for details and further discussion).

\begin{figure}
	\begin{center} 
		\includegraphics[width=\textwidth]{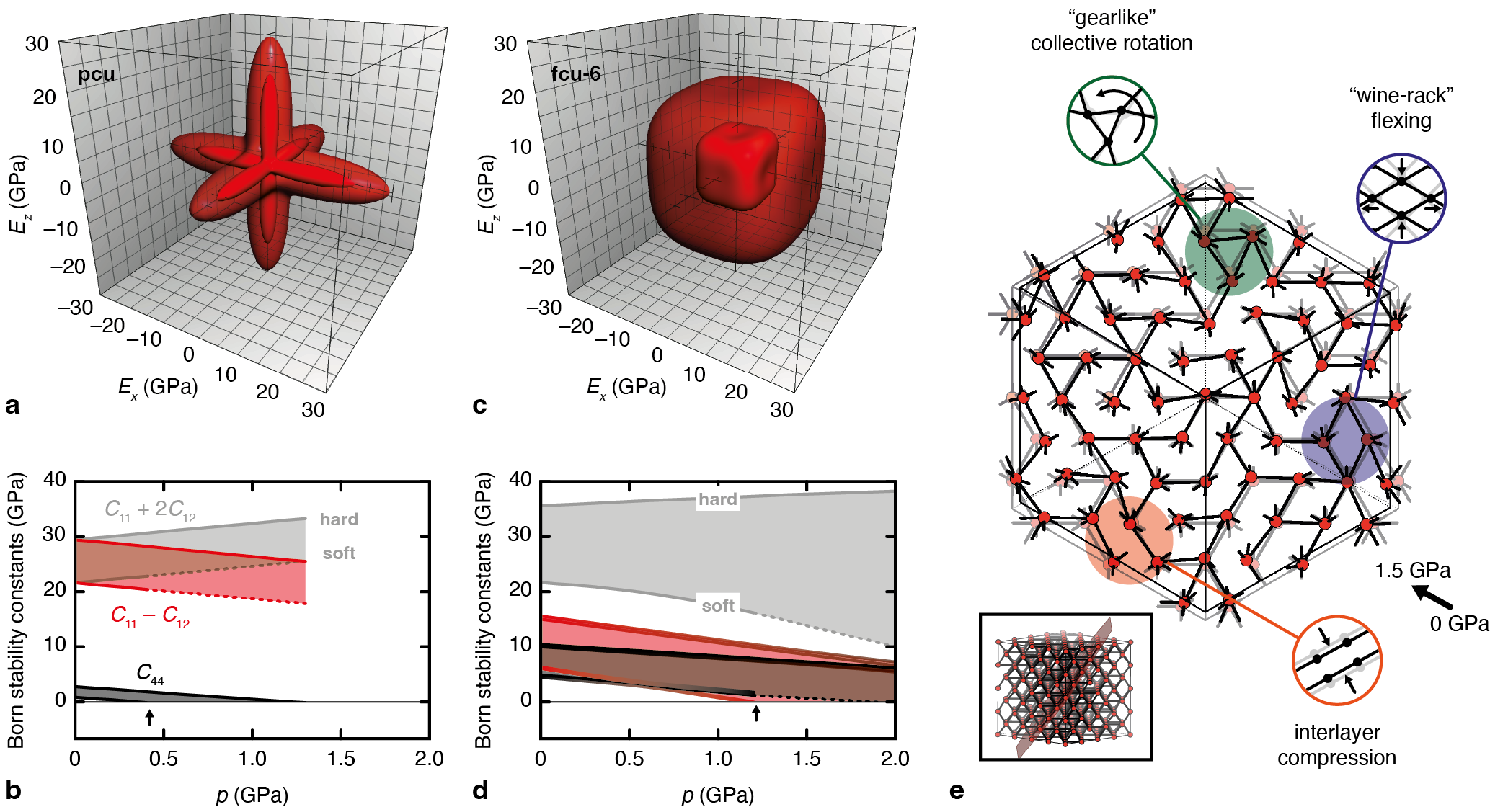}
	\end{center}
	\caption{\label{fig3}\footnotesize{\bf Elastic behaviour of coarse-grained MOF-5 and TRUMOF-1 models.} {\bf a} The orientational dependence of the Young's modulus calculated from Eq.~\eqref{hamil1} applied to the {\bf pcu} topology of MOF-5, parameterised as discussed in the text. The `soft' (`hard') parameter set gives rise to the interior (exterior) surface of the indicatrix as drawn. {\bf b} The pressure-dependence of the Born stability constants for this {\bf pcu} model, defined as the eigenvalues of the elastic stiffness tensor $\mathbf B$ under load.\cite{Cowley_1976,Mouhat_2014} Shaded regions correspond to the values spanned by the `soft' and `hard' parameterisations discussed in the text; the onset of elastic instability is marked by an arrow. {\bf c} The orientational Young's modulus surfaces for the coarse-grained TRUMOF-1 structure with {\bf fcu-6} topology, shown using the same representation used in (a) for MOF-5. {\bf d} Pressure dependence of the Born stability constants for our coarse-grained TRUMOF-1 model, revealing an enhanced elastic stability relative to MOF-5. Formally, the absence of crystallographic symmetry in the approximant structure allows arbitrary mixing of the eigenstates of the elastic tensor. We have coloured the stability constants by similarity to the $C_{11}+2C_{12}, C_{11}-C_{12}, C_{44}$ eigenstates shown in {\bf b}. {\bf e} Representation of characteristic node displacements induced by compression of TRUMOF-1, shown here for a $(111)$ slice of a $4\times4\times4$ approximant (inset). Compression mechanisms vary spatially, and some easily-interpreted mechanical motifs are highlighted by coloured circles. Dashed lines in {\bf b} and {\bf d} denote extrapolations beyond the corresponding elastic stability regime.} 
\end{figure}

Irrespective of the particular parameterisation employed, the elastic behaviour described by Eq.~\eqref{hamil1} captures qualitatively the key response of MOF-5 to external stress. For example, the orientational dependence of the Young's modulus\cite{Gaillac_2016} is similar to that identified in earlier studies, in terms of both degree and form of anisotropy [Fig.~\ref{fig3}(a)].\cite{Tan_2011,Banlusan_2017,Rogge_2018} Likewise, as pressure is increased, the $C_{44}$ elastic constant softens, with the system becoming mechanically unstable beyond $p\simeq0.3$\,GPa [Fig.~\ref{fig3}(b)]. Since the connectivity in \eqref{hamil1} is fixed, this simple model cannot reproduce amorphisation; however, the shear instability of MOF-5 is considered a key element in driving amorphisation in practice, and it also places an upper bound on the pressure stability of the material.\cite{Mouhat_2014,Bhogra_2021}

Our next step was to transpose this coarse-grained model to TRUMOF-1, retaining the same range of effective spring constants $k_r$ and $k_\theta$ used for MOF-5 (since the chemistries are so similar), and using the aperiodic \textbf{fcu-6} network connectivity identified in Ref.~\citenum{Meekel_2023} to assign neighbour pairs and triplets amongst which these harmonic interactions operate. The equilibrium value $r_{\rm e}$ was set to the node--node separation of the TRUMOF-1 structure ($=a/\sqrt{2}$), and the equilibrium angles $\theta_{\rm e}$---now a function of triplet $i,j,k$---were set to the corresponding values when nodes $i,j,k$ were placed at the high-symmetry (face-centred) sites in the configurationally-averaged $F\bar43m$ TRUMOF-1 cell. In this way, the ground state of Eq.~\eqref{hamil1}, when applied to TRUMOF-1, corresponds to a face-centred cubic arrangement of nodes, as observed in the experimental average structure. Our use of a coarse-grained model allows interrogation of much larger approximants to the aperiodic structure of TRUMOF-1 than are accessible using \emph{e.g.}\ DFT. 

The elastic behaviour of this simplified TRUMOF-1 model turns out to be interesting in a number of respects. First, the Young's modulus is now very isotropic [Fig.~\ref{fig3}(c)]: using the metric of Ref.~\citenum{Gaillac_2016} we obtain an elastic anisotropy $1.3<E_{\rm max}/E_{\rm min}<1.5$ much reduced from that of MOF-5 ($4.0<E_{\rm max}/E_{\rm min}<9.3$). Indeed the form of the Young's modulus we obtain is qualitatively similar to those of the two isotropic periodic elastic networks (\emph{viz}.\ {\bf crs} and {\bf hxg}) derivable from the {\bf fcu} topology by removing half of the elastic links (see SI for further discussion). Second, the system remains elastically stable well beyond 1\,GPa [Fig.~\ref{fig3}(d)]---in other words, the enhanced elastic stability observed experimentally is successfully reproduced by this coarse-grained model. Moreover, the absence of any crystallographic symmetry ensures that the system is dynamically stable (\emph{i.e.}\ real-valued phonon frequencies) throughout this same pressure range.\cite{Bhogra_2023} And, third, the $V(p)$ equation of state gives a low-pressure bulk modulus $7.2<B_0<12$\,GPa that is remarkably close to the experimental value [Fig.~\ref{fig1}(a)]. The only qualitative feature of $V(p)$ not captured by the coarse-grained model is the curvature at high pressure, which is perhaps unsurprising given the omission of any non-bonded interactions. Nevertheless we conclude that the contrasting elastic stabilities of MOF-5 and TRUMOF-1 can indeed be rationalised in terms of the difference in the topologies of their underlying elastic networks.

Whereas in MOF-5 the relative arrangements of nodes within the unit cell is unaffected by pressure (their positions being fixed by crystal symmetry), the equilibrium node positions in TRUMOF-1 are free to vary as pressure is applied. The spatially-varying network connectivity then results in an increased displacement of nodes away from their high-symmetry sites at elevated pressures [Fig.~\ref{fig3}(e)].\cite{DFTnote} One expects a signature of this behaviour in the crystallographic model, which represents a configurational average over all unit cells. We extracted from our lattice-dynamical calculations the pressure dependence of the mean-squared node displacements $U(p)$, which is a measure of the additional static disorder introduced by spatial variations in the elastic network of TRUMOF-1. This function maps closely onto the increase in magnitude of atomic displacements measured experimentally [Fig.~\ref{fig2}(b)]. Our DFT configurations, which necessarily involve smaller unit cells with higher crystal symmetry, respond to pressure through a combination of internal reorganisation and increasing anisotropic strain; we find the latter dominates but this contribution can nonetheless be recast in terms of a static disorder contribution, which again follows experiment relatively closely (Fig.~\ref{fig2}(b); see SI for further discussion). Hence we attribute the anomalous increase in the value of $U$ with pressure to an increase in the degree of static disorder of the underlying TRUMOF-1 network, itself a consequence of topological aperiodicity. For context, we note that a related effect has been observed previously in the low-temperature crystallography of  various MOFs, where spatial inhomogeneities in host--guest interactions induced by cooling result in an anomalous increase in $U$ with decreasing $T$.\cite{Collings_2012,Lee_2018}

\section*{Discussion}

The node displacement patterns that emerge from our simple lattice-dynamical calculations provide further qualitative insight into the pressure-induced mechanical response of TRUMOF-1. Macroscopic strain is accommodated through large-scale node displacements that, although correlated between connected nodes, are nonetheless spatially localised and do not propagate cooperatively. This localisation may act to frustrate collective instabilities in a manner analogous to that proposed elsewhere for stiffening disordered metamaterials.\cite{Liu_2022} An interesting feature of the distortion pattern that emerges, as shown in Fig.~\ref{fig3}(e), is that different mechanisms---\emph{e.g.}\ wine-rack flexing,\cite{Ortiz_2013} layer compression,\cite{Adamson_2015} and even gearlike motion\cite{Duyker_2016}---are activated in different regions of the aperiodic network. That disordered networks might provide a material with a library of distinct elastic response mechanisms that can be amplified selectively so as to best accommodate a given strain is a concept termed ``combinatorial mechanics''---a mechanical analogue of the response of dynamic combinatorial libraries to chemical perturbation.\cite{Lehn_2001,Reynolds_2021} Whereas in randomly diluted elastic networks, bulk elasticity is dominated by the existence (or absence) of a single percolating cluster,\cite{Jacobs_1995,Jacobs_1996} the uniform local connectivity in TRUMOF-1 appears to drive a more complex and spatially distributed response.

Given the strong coupling between local elastic connectivity and the corresponding mechanical response of the {\bf fcu-6} network, one expects fundamental differences in the vibrational properties of TRUMOF-1 relative to those of conventional periodic network structures. In particular, the spatial localisation of different response mechanisms suggests a breakdown of the usual phonon description for this material, with mode broadening in both wave-vector and energy.\cite{Roth_2023} Such behaviour is known from theory to arise in related systems based on strongly correlated disorder,\cite{Thebaud_2023} where the interest is in exploiting anomalous phonon broadening as a design principle for inhibiting thermal transport in thermoelectrics. Hence the anomalous mechanical response of TRUMOF-1 to hydrostatic pressure (as we report here) may also provide insight in due course into the dynamical response of the same material as a function of temperature. We anticipate that Truchet-tile architectures may allow engineering of anomalous thermal responses in precisely the same way that we have now found TRUMOF-1 to exhibit a combination of elastic compliance and elastic stability not observed in conventional, crystalline MOFs.

\section*{Acknowledgments}
A.L.G. gratefully acknowledges M. T. Dove (QMUL) for useful discussions. The authors acknowledge financial support from the E.R.C. (Advanced Grant 788144 to A.L.G.) the Royal Society (Industry Fellowship Grant IF160062 to B.S.), the UKRI (Future Leaders Fellowship MR/V026070/1 to C.L.H.) and the University of Edinburgh (Chancellor's Fellowship to C.L.H.), and acknowledge further support in the form of access to the UK’s ARCHER2 supercomputer through the Materials Chemistry Consortium, which is funded by the Engineering and Physical Sciences Research Council (Grants EP/R029431 and EP/X035859). The authors gratefully acknowledge the provision of synchrotron beamtime on the I19 Beamline at the Diamond Light Source, Harwell, U.K.

\section*{Data Availability}
The authors declare that the data supporting the findings of this study are available within the paper and its supplementary information files.

\section*{References and Notes}
\bibliography{nmat_2023_trumofhighp}

\begin{thebibliography}{10}
\bibitem{DFTnote}
\bibinfo{title}{Note that our {DFT} calculations were performed on the densest
  $1\times1\times1$ approximant structure, the connectivity of which has
  {${P}\bar4$} crystal symmetry and so constrains the relative coordinates of
  cluster positions. {Consequently} we are unable to meaningfully compare the
  static disorder introduced in the {DFT} configurations with our that
  extracted from either experiment or our coarse-grained model.}
\end{thebibliography}

\end{document}